# ScarNet: A Novel Foundation Model for Automated Myocardial Scar Quantification from LGE in Cardiac MRI

Neda Tavakoli[a], Amir Ali Rahsepar[a], Brandon C. Benefield[b], Daming Shen[c], Santiago López-Tapia[d], Florian Schiffers[d], Jeffrey J. Goldberger[e], Christine M. Albert[f], Edwin Wu[b], Aggelos K. Katsaggelos[d], Daniel C. Lee[b], Daniel Kim[a]

[a] Department of Radiology, Feinberg School of Medicine, Northwestern University, Chicago, IL, USA.

[b] Department of Medicine (Division of Cardiology), Northwestern University Feinberg School of Medicine, Chicago, IL, USA.

[c] Department of Biomedical Engineering, Northwestern University, Evanston, IL, USA.

[d] Department of Electrical and Computer Engineering, Northwestern University, Evanston, IL, USA.

[e] Cardiovascular Division, University of Miami Miller School of Medicine, Miami, FL, USA.

[f] Smidt Heart Institute, Cedars-Sinai Medical Center, Los Angeles, CA, USA.

Corresponding Author: Neda Tavakoli, PhD
Department of Radiology, Feinberg School of Medicine, Northwestern University, 737 North Michigan Avenue, Chicago, Illinois 60611, USA., Twitter: @neda_tv, Email Address: **neda.tavakoli@northwestern.edu**

**Grant Support:** The National Institutes of Health (R01HL116895, R01HL151079, R01HL167148, R01HL168859), the Radiological Society of North America (EILTC2302), and the American Heart Association (949899).

**Abstract**

**Background:** Late Gadolinium Enhancement (LGE) imaging remains the gold standard for assessing myocardial fibrosis and scarring, with left ventricular (LV) LGE presence and extent serving as a predictor of major adverse cardiac events (MACE). Despite its clinical significance, LGE-based LV scar quantification is not used routinely due to the labor-intensive manual segmentation and substantial inter-observer variability.

**Methods:** We introduce ScarNet that synergistically combines a transformer-based encoder in Medical Segment Anything Model (MedSAM) and a convolution-based decoder in U-Net with tailored attention blocks to automatically segment myocardial scar boundaries while maintaining anatomical context. This network was trained an existing database of 552 ischemic cardiomyopathy patients with expert segmentation of myocardial and scar boundaries in LGE images, and it was tested on a separate set of 184 patients.

**Results:** In 184 testing (not seen during training) patients, ScarNet achieved accurate segmentation of the scar boundary (median DICE= 0.912 [IQR: 0.863-0.944]), significantly outperforming both MedSAM (median DICE= 0.046 [IQR: 0.043-0.047]) and nnU-Net (median DICE= 0.638 [IQR: 0.604-0.661]). The bias and the coefficient of variation (CoV) were considerably lower for ScarNet (difference = -0.63% [4.82% of mean]; CoV = 4.3%) than MedSAM (difference = -13.31% [198.18% of mean]; CoV = 130.3%) and nnU-Net (difference = -2.46% [20.31% of mean]; CoV = 20.3%). In the Monte Carlo simulation with noise perturbation, ScarNet (0.892 ± 0.053; CoV = 5.9%) produced significantly higher scar DICE than MedSAM (0.048 ± 0.112; CoV = 233.3%) and nnU-Net (0.615 ± 0.537; CoV = 28.7%).

**Conclusion:** ScarNet outperformed MedSAM and nnU-Net for predicting myocardial and scar boundaries in LGE images of patients with ischemic cardiomyopathy. Using Monte Carlo simulation, the model demonstrated robust performance across varying image qualities and scar patterns.

## INTRODUCTION

Cardiovascular disease remains a leading cause of mortality worldwide, accounting for approximately 32% of all global deaths [1]. Myocardial injury frequently results in myocardial scarring in both ischemic and non-ischemic etiologies [2]. Late Gadolinium Enhancement (LGE) cardiovascular magnetic Resonance (CMR) [3] is the gold standard non-invasive test for myocardial fibrosis and scarring, providing essential information for risk stratification, treatment planning, and prognosis evaluation. Among LGE metrics, left ventricular (LV) scar volume is especially significant, as it has proven to be a predictor of major adverse cardiac events (MACE) and arrhythmic complications [4].

Despite its clinical importance, LGE-based LV scar quantification remains challenging to implement in practice due to several factors [5]. Traditional manual segmentation requires around 15-20 minutes per case and is highly user-dependent, with considerable inter-observer variability [6-8]. This variability stems from the complex and heterogeneous appearance of scar tissue in LGE images, which often present with diffuse boundaries, irregular shapes, and varying intensity patterns. Additionally, imaging artifacts such as motion blur, intensity inhomogeneity, ghosting artifacts arising from arrhythmia and noise further complicate accurate scar delineation [9], making reliable manual segmentation both labor-intensive and inconsistent.

Recent advancements in deep learning (DL) have inspired efforts to automate LV scar segmentation in LGE [10]. Traditional convolutional neural networks (CNNs), particularly UNet-based architectures [11, 12], have shown promise in medical image segmentation tasks. However, these methods have encountered limitations in cardiac scar segmentation. A 2021 meta-analysis of 35 AI-based left ventricular scar quantification studies demonstrated mean DICE scores of only 0.616 and 0.633 for supervised and unsupervised methods, respectively [13]. These limited performances are attributed to the inherent challenges of capturing long-range dependencies in cardiac anatomy, handling multi-scale variations in scar appearance [14], and accurately delineating poorly defined scar boundaries.

The recent development of transformer-based models [15], including the Segment Anything Model (SAM) [15, 16] and its medical variant, MedSAM [17], has introduced new possibilities for medical image segmentation. These models excel at capturing global context and handling diverse object appearances, presenting a unique opportunity to address the challenges of cardiac scar segmentation. However, their generalist nature and lack of domain-specific optimization limit their direct applicability to LGE analysis [18, 19].

To address these challenges, we propose ScarNet, a novel foundation model designed specifically for LV scar segmentation in LGE as illustrated in Fig. 1. ScarNet integrates the global context-awareness of MedSAM's transformer-based encoder with the precise localization capabilities of a UNet decoder through an adaptive fusion mechanism. This architecture is further strengthened by specialized attention mechanisms, including multi-scale feature extraction pathways enhanced by Squeeze-and-Excitation (SE) [20] modules and novel ScarAttention blocks that dynamically focus on subtle scar regions while preserving anatomical context.

To address the class imbalance inherent in cardiac scar segmentation, ScarNet incorporates a comprehensive multi-component loss function. This loss function combines DICE, Focal [21], and Cross-Entropy losses with class-specific optimization terms for scar and myocardium segmentation, supported by an adaptive weighting scheme to balance the contributions of each component throughout training.

Our novel ScarNet architecture addresses key challenges in cardiac scar segmentation through several innovative design elements. The hybrid structure combines global context understanding with precise local feature detection, while specialized attention mechanisms enhance the capture of heterogenous scar patterns. Furthermore, an efficient inference pipeline supports seamless integration into clinical workflows, and the adaptive fusion design enables comprehensive multi-scale feature extraction. The main contributions of our work are as follows:

1. We propose ScarNet, a unique hybrid architecture that fine-tunes and extends MedSAM's transformer-based encoder by combining it with a UNet decoder. This adaptive fusion mechanism leverages both the global context awareness of transformers and the precise localization capabilities of CNNs, enabling comprehensive feature extraction at multiple scales.
2. ScarNet incorporates multi-scale feature extraction pathways enhanced by SE modules for refined feature representation. The model includes ScarAttention blocks that dynamically focus on subtle scar features while preserving anatomical context and applies an adaptive fusion strategy to optimally combine features from both MedSAM and UNet streams.
3. We propose a multi-component loss function that includes a weighted combination of DICE, Focal, and Cross-Entropy losses. This loss function integrates class-specific optimization terms for scar and myocardium segmentation, as well as an adaptive weighting scheme to balance different loss components during training.
4. We demonstrate ScarNet's superior performance, achieving minimal volume quantification error (−4.82%) compared to MedSAM and nnU-Net. The model shows robust generalization across varied scar patterns and imaging conditions, along with efficient inference times that support seamless integration into clinical workflows.
5. This work establishes a novel framework that combines transformer-based and CNN architectures for medical image segmentation, with potential applications extending beyond cardiac imaging.

The purpose of this study was to implement ScarNet and compare its performance against MedSAM and nnU-Net on an existing database of LGE images of ischemic cardiomyopathy patients with expert segmentation of myocardial and scar boundaries as reference.

## METHODS

### Patient Cohort

This is a retrospective study using de-identified LGE images derived from 736 patients (mean age = 57.82 ± 18.58 years; 60% males, mean left ventricular ejection fraction = 40.3±11.0%) with coronary artery disease and left ventricular dysfunction enrolled in the previous DETERMINE registry (ClinicalTrials.gov ID NCT00487279) and PRE-DETERMINE study (ClinicalTrials.gov ID NCT01114269) [22] (see Table 1 for participant demographics and characteristics). Additional clinical profiles were not available to the study team at the time of this study. LGE images were manually segmented using the full width at half maximum technique [23] in QMass software (version 7.6, Medis Medical Imaging Systems, Leiden), by two cardiology attendings with more than 10 years of clinical experience in CMR. For additional details of the dataset and data preparation, please see Appendix in the Supplementary Materials.

### Model Architecture

We include a literature review of related works to highlight their relative strengths and weaknesses, ultimately justifying the need for our study. Please see Appendix in Supplementary Materials for this literature review.

The ScarNet architecture, shown in Fig. 1 and Fig. 2, integrates global and local features by combining the pretrained MedSAM with a UNet pathway via an adaptive fusion mechanism, effectively capturing both anatomical context and spatial details necessary for precise LV scar segmentation in LGE. The MedSAM pathway uses a Vision Transformer (ViT) backbone, pretrained on large-scale general and medical datasets, to leverage global contextual information and subtle structural relationships in cardiac imaging. The complete segmentation pipeline is detailed in Algorithm 1, which outlines the key steps of our approach from input processing to final mask generation. The feature fusion process between MedSAM and UNet pathways is elaborated in Algorithm 4. Additional algorithmic details can be found in the Appendix.

By dividing the input image into non-overlapping patches, each mapped into an embedding space, MedSAM processes these features through a multi-scale decoder that progressively reduces channel dimensions from 256 to 32, preserving essential spatial relationships. This pathway is further enhanced with Squeeze-and-Excitation (SE) blocks that recalibrate channels, highlighting scar-relevant regions while suppressing background features, thus maximizing the representational power of the pretrained MedSAM (Eq. 3–4).

In parallel, the U-Net pathway captures localized spatial details through a symmetric encoder-decoder architecture, progressively reducing spatial resolution while increasing feature depth with each encoding block (Eq. 5). Skip connections preserve high-resolution features by linking corresponding encoder and decoder blocks, which enables accurate boundary delineation critical for scar segmentation (Eq. 6). The adaptive fusion mechanism combines outputs from both pathways, balancing global and local context by aligning feature maps and applying an adaptive weighting mechanism. This fusion consolidates ScarNet's learned representations into a single four-class segmentation map—distinguishing background, myocardium, blood pool, and scar regions—through a final 1×1 convolutional layer.

#### A. Attention Mechanisms

ScarNet's attention mechanisms focus on scar-relevant features, enhancing the model's ability to generalize across diverse imaging conditions. SE modules, embedded within the MedSAM pathway, recalibrate channel-wise features by modeling interdependencies to boost sensitivity to scar tissue regions, while suppressing irrelevant background areas. Complementing this, the ScarAttention blocks, detailed in Algorithm 2, dynamically emphasize scar-specific features by reweighting spatial attention. These blocks use positional relationships and intensity variations unique to scars, improving the model's accuracy in highlighting subtle scar regions while preserving surrounding anatomical structures.

### B. Training Pipeline

ScarNet's training pipeline, described in Algorithm 1, is structured to ensure optimal segmentation capabilities. Initially, input LGE images are normalized and fed into both MedSAM and UNet pathways. The parallel processing of these pathways generates both global and local feature representations, which are then enhanced by SE and ScarAttention modules before being fused to yield a four-class segmentation map. Loss calculation follows, with $\mathcal{L}_{ScarNet}$ computed based on the adaptive weighting of DICE, Focal, and Cross-Entropy losses, guiding the model to optimize for both global and local accuracy. Backpropagation is conducted using an Adam optimizer, with gradients calculated to minimize $\mathcal{L}_{ScarNet}$ (Algorithm 1, Steps 13-15). Regularization techniques, such as dropout and data augmentation, are applied to improve generalization.

During inference, ScarNet processes unseen LGE images and generates segmentation maps for background, myocardium, blood pool, and scar regions, with the inference algorithm detailed in Algorithm 4 dynamically adjusting feature weighting for efficient predictions.

To assess ScarNet's robustness and reliability, we conducted a Monte Carlo simulation using 200 iterations with a Gaussian noise level of 0.05. This noise level was chosen to simulate realistic image quality variations encountered in clinical settings while providing a test of model stability. Each iteration introduced random perturbations to the input images while maintaining consistent noise patterns through controlled random seed initialization, allowing for reproducible yet comprehensive stress testing of the model's performance.

### C. Mathematical model

The ScarNet model, denoted by $\mathcal{H}$, combines a fine-tuned and enhanced MedSAM and UNet architecture to achieve accurate segmentation of cardiac scar regions. The model's core function can be expressed as:

$$\mathcal{H}(\mathrm{x}) = \mathcal{F}\big(\mathrm{MedSAM}(\mathrm{x}), \mathrm{UNet}(\mathrm{x})\big) \quad (1),$$

where $x \in R^{H\times W\times 1}$ represents a grayscale medical image, and $\mathcal{F}$ is a fusion function that combines the outputs of the MedSAM and UNet branches. The fusion function $\mathcal{F}$ takes two feature maps, $f_1$ and $f_2$, produced by MedSAM and UNet, respectively, and combines them through concatenation and a 1x1 convolution to produce the final segmentation output:

$$\mathcal{F}(\mathrm{f}_1, \mathrm{f}_2) = \mathrm{Conv}(\mathrm{Concat}[\mathrm{f}_1, \mathrm{f}_2]) \quad (2),$$

where Conv is a 1x1 convolution that reduces the concatenated feature map to the desired number of classes, and Concat represents channel-wise concatenation.

*1) MedSAM Pathway*

The MedSAM component of ScarNet leverages a ViT backbone to capture global dependencies within the image, which is crucial for identifying subtle structures like scar tissue. MedSAM consists of patch embedding and position encoding stages, where the input image $\mathrm{x} \in \mathrm{R}^{\mathrm{H}\times\mathrm{W}\times 1}$ is divided into $\mathrm{p} \times \mathrm{p}$ non-overlapping patches. Each patch is flattened and mapped to an embedding space, generating patch embeddings $E$:

$$\mathrm{E} = \mathrm{W}_{\mathrm{embed}} \cdot \mathrm{P}(\mathrm{x}) \quad (3),$$

where $P(x)$ denotes the patches, and $W_{\text{embed}}$ is a learnable embedding matrix. The transformer operates on these patch embeddings, applying self-attention to learn relationships across the entire spatial dimension of the image. For query Q, key K, and value V matrices derived from E, self-attention is computed as:

$$Attention(Q,K,V) = softmax\left(\frac{QK^T}{\sqrt{d}}\right)V \quad (4),$$

where d is the embedding dimension, enabling global context learning. The encoded features are passed through multiple Attention Blocks and SE layers in a multi-scale decoder. This process enhances the features relevant to scar regions. The detailed processing steps of the MedSAM branch are presented in Algorithm 2, highlighting the transformer-based processing and attention mechanisms.

*2) U-Net Pathway*

The U-Net pathway captures local spatial details via a hierarchical encoder-decoder structure. The encoder consists of convolutional layers followed by downsampling steps, progressively reducing spatial resolution and increasing feature channels:

$$F_{enc}^{l} = Downsample\left(f_{enc}^{l-1}\right) \quad (5).$$

Each level l extracts features at a smaller spatial scale, enabling the capture of localized features. The decoder restores spatial resolution by upsampling, integrating high-resolution features from the encoder via skip connections:

$$F_{dec}^{l} = Upsample\left(f_{dec}^{l+1}\right) + f_{enc}^{l} \quad (6).$$

The final layer produces a feature map $M_{\text{UNet}}$in $R^{H\times W\times C}$, where C is the number of classes. Algorithm 3 provides a detailed breakdown of the U-Net branch operations, including the encoder-decoder pathway and skip connection enhancement steps.

*1) Loss function*

In ScarNet, the overall loss function $\mathcal{L}_{ScarNet}$ is designed to address class imbalance and to focus on achieving high segmentation accuracy, especially for challenging regions like cardiac scars. The ScarNet model aims to minimize the overall loss function with respect to the model parameters θ, ensuring accurate segmentation across classes, with a particular emphasis on the scar class. This can be formulated as:

$$\theta^* = \arg\min_{\theta} \mathcal{L}_{\text{ScarNet}}(\theta) \quad (7),$$

where $\theta^*$ represents the optimal parameters of the model, and $\mathcal{L}_{ScarNet}$ is the combined loss function defined as:

$$\mathcal{L}_{ScarNet} = \lambda_1.FTL + \lambda_2.DL + \lambda_3.CE \quad (8),$$

where, $FTL$ is the Focal Tversky Loss, which mitigates class imbalance by focusing more on the scar region $DL$ is the DICE Loss, which measures spatial overlap between the predicted and ground truth masks, $CE$ is the Cross-Entropy Loss, weighted by class-specific importance $\lambda_1$, $\lambda_2$, and $\lambda_3$ are coefficients that control the contribution of each term, tuned to prioritize scar segmentation. The optimization is subject to constraints that guide the model's focus:

1. Class Imbalance Constraint: Emphasis is placed on the scar class by weighting the Focal Tversky Loss term $FTL$

2. more heavily, enabling the model to manage class imbalances effectively. This constraint prioritizes reducing false negatives and false positives in scar segmentation, thus emphasizing the critical region of interest.
3. Spatial Accuracy Constraint: The DICE Loss $DL$ component ensures that the segmentation output maintains spatial alignment with the ground truth by maximizing overlap. This constraint encourages accurate boundary delineation, especially for small or complex structures.
4. The Cross-Entropy Loss $CE$, with class-specific weights, further reinforces focus on critical regions, allowing flexibility in segmentation where non-scar regions may have less impact on the overall segmentation goal.

Focal Tversky Loss (FTL) is adapted from the Tversky Index, emphasizing false positives and false negatives to handle class imbalance effectively. FTL is defined as:

$$\text{FTL} = \left(1 - \frac{\sum(p \cdot g) + \epsilon}{\sum(p \cdot g) + \alpha \sum\big(p \cdot (1-g)\big) + \beta \sum\big((1-p).g\big) + \epsilon}\right)^{\gamma} \quad (9).$$

DICE Loss helps maximize the overlap between predicted $p$ and ground truth $g$, masks, improving spatial accuracy. The DL is calculated as:

$$DL = 1 - \frac{2 \cdot |\text{p} \cap \text{g}| + \epsilon}{|\text{p}| + |\text{g}| + \epsilon} \quad (10).$$

Cross-Entropy Loss is calculated for each pixel, weighted by class to improve representation of the scar class over background regions:

$$\text{CE} = -\sum_{c} w_c \cdot g_c \log(p_c) \quad (11),$$

where $w_c$ is the weight for class $c$ and $p_c$ and $g_c$ are the predicted and true probabilities for class c, respectively.

This loss formulation provides a balanced objective that aligns ScarNet to achieve high accuracy in cardiac scar segmentation by reducing errors across both global context (captured by MedSAM) and local detail (captured by U-Net). The training process aims to minimize $\mathcal{L}_{ScarNet}$ by updating the model weights to reduce misclassifications, particularly for the scar class. This is achieved by Gradient Descent Optimization: Using optimizers like Adam, gradients of $\mathcal{L}_{ScarNet}$ with respect to model parameters are computed. These gradients indicate the direction and magnitude of updates needed to minimize the loss. Regularization methods like dropout (if applicable) and data augmentation help the model generalize, while tuning the weights $\lambda_1$, $\lambda_2$, and $\lambda_3$ in the combined loss function focuses the model on scar segmentation without overfitting.

ScarNet demonstrates superior performance in cardiac scar segmentation by effectively delineating scar boundaries, reducing false positives, and providing consistent segmentation quality across varying scar sizes, shapes, and image qualities. By combining MedSAM's ability to capture global context with U-Net's precision in local feature extraction, ScarNet creates a robust foundation for accurate cardiac scar segmentation. The use of attention mechanisms at multiple scales further enhances the model's ability to focus on relevant features, resulting in reliable and precise segmentation outcomes.

Additional implementation details of our network are available in Supplementary Materials.

### *Two Secondary Experiments to Demonstrate Robustness*

First, to compare the performance of our model against other models as a function of training data size, we trained and tested the models with training-testing (maintaining a 0.25 ratio) size ranging from 15 to 552 patients. Second, to compare the performance of different models against noise perturbation, we conducted a Monte Carlo simulation using identical testing conditions (200 iterations, noise level 5%). This noise level was empirically determined to be an upper limit in clinical practice by a cardiothoracic Radiologist attending with 7 years of clinical experience in CMR.

### *Statistical Analyses*

We tested for variable normality using the Kolmogorov-Smirnov, Anderson-Darling, and Shapiro-Wilk tests. A variable was deemed normality distributed if it passes all three tests. Bland-Altman analysis was performed on DICE scores to assess the level of agreement between measurements, and the coefficient of variation (CoV) was defined as the standard deviation of the difference divided by the mean. One-way analysis of variation (Kruskal-Wallis if not normally distributed) with Bonferroni correction was conducted to detect any significant differences among groups. A p-value $< 0.05$ was considered statistically significant for all tests performed.

## *Results*

Figure 3 shows a comparison of segmentation performance between MedSAM and ScarNet and their corresponding feature logits and probability maps. This figure does not include nnU-Net, because it does not provide feature maps. As shown, ScarNet focused on the correct anatomic features better than MedSAM for predicting the region of interests.

Figure 4 compares the performance of MedSAM, nnU-Net, and ScarNet as a function of training data size. For both myocardium and scar segmentations, ScarNet achieved higher accuracy throughout and hit the plateau faster. As shown in the violin plots combing all training data sizes in Fig. 4, ScarNet not only produced higher median scar DICE (0.912) but also tighter score distributions compared to MedSAM (0.128) and nnU-Net (0.375) with lower median DICE scores and considerably wider distributions. Likewise, the same trends were observed for the myocardial DICE scores.

The remaining results are from the training size of 552 patients and testing size of 184 patients. Figure 5 compares segmentation performance between MedSAM, nnU-Net, and ScarNet in four representative patients, where ScarNet consistently outperformed the other networks. The DICE scores and scar volumes in 184 testing patients were not normally distributed ($p < 0.05$), necessitating non-parametric analyses. Because ScarNet produced the highest DICE scores, we compared differences with ScarNet as the reference. Figure 6 shows statistical results: for myocardial segmentation, compared against ScarNet (median DICE=0.961 [IQR: 0.920-0.999]), nnU-Net (median DICE= 0.878 [IQR: 0.838-0.915]) was significantly ($p<0.001$) different, whereas MedSAM (median DICE = 0.242 [IQR: 0.116-0.342]) was also significantly ($p<0.001$) different. For the scar segmentation, compared with ScarNet (median DICE= 0.912 [IQR: 0.863-0.944]), MedSAM (median DICE= 0.046 [IQR: 0.043-0.047]) and nnU-Net (median DICE= 0.638 [IQR: 0.604-0.661]) were significantly ($p<0.001$) different. For scar volume quantification, compared against manual as the reference (median DICE= 0.114 [IQR: 0.094-0.180]), MedSAM (median DICE= 0.000 [IQR: 0.000-0.001]) was significantly ($p<0.001$) different, whereas ScarNet (median DICE = 0.109 [IQR: 0.087-0.173]) and nnU-Net (median DICE= 0.095 [IQR: 0.068-0.150]) were not significantly ($p>0.192$ and $p>0.0009$, respectively) different. According to the Bland-Altman analysis with manual scar volume as the reference (Figure 7), the bias and the CoV were considerably lower for ScarNet (difference = -0.63% [4.82% of mean]; CoV = 4.3%) than MedSAM (difference = -13.31% [198.8% of mean]; CoV = 130.3%) and nnU-Net (difference = -2.46% [20.31% of mean]; CoV = 20.31%). Correlation analysis of scar volume measurements further supported these findings, with ScarNet showing near-perfect correlation with manual measurements ($R^2$ = 1.0, y = 0.96x - 0.11), while nnU-Net demonstrated good correlation ($R^2$ = 0.94, y = 0.8x + 0.21), and MedSAM showed poor correlation ($R^2$ = 0.65, y = 0x + 0.01).

Figure 8 shows representative examples with noise added in the Monte Carlo simulation. The mean of average myocardial DICE for ScarNet (0.912 ± 0.063) was significantly ($p<0.001$) higher than MedSAM (0.185 ± 0.142) and nnU-Net (0.823 ± 0.072), and the mean of average scar DICE for ScarNet (0.892 ± 0.053) was significantly ($p<0.001$) higher than MedSAM (0.048 ± 0.112) and nnU-Net (0.615 ± 0.537). Likewise, the mean CoV of myocardial DICE for ScarNet (6.9%) was significantly ($p<0.001$) higher than MedSAM (76.8%) and nnU-Net (8.7%), and the mean CoV of scar DICE for ScarNet (5.9%) was significantly ($p<0.001$) higher than MedSAM (233.3%) and nnU-Net (28.7%).

## Discussion

In this study, we developed ScarNet, a novel deep learning model that combines a complete MedSAM pathway with a parallel UNet pathway through an adaptive fusion mechanism for automated myocardial scar segmentation in LGE images. After fine-tuning the model using a dataset of 552 patients and evaluating on 184 test patients, ScarNet achieved superior segmentation accuracy for both myocardium (median DICE=0.961) and scar tissue (median DICE=0.912), significantly outperforming both MedSAM and nnU-Net. Using Monte Carlo simulations with 5% Gaussian noise perturbation, the model demonstrated exceptional robustness with minimal bias (-0.63%) and coefficient of variation (4.3%) in scar volume quantification, while maintaining consistently high DICE scores for both myocardium (0.912 ± 0.063; CoV = 6.9%) and scar tissue (0.892 ± 0.053; CoV = 5.9%).

Our approach differs from previous deep learning methods by addressing several key limitations in automated scar segmentation. Traditional CNN-based approaches [10-12], while effective for general medical segmentation, struggle with the inherent challenges of LGE imaging, achieving limited DICE scores of only 0.616 and 0.633 for supervised and unsupervised methods, respectively [13]. Earlier efforts by Zabihollahy et al. [24] and Bai et al. [25] demonstrated the potential of CNNs for scar segmentation, but faced challenges in capturing complex scar patterns. A significant advancement came from Fahmy et al., who introduced a deep learning-based image fusion approach that improved scar quantification accuracy by combining multiple image features [12]. While their method showed promise in handling varying contrast patterns, it still faced challenges with complex scar morphologies and required careful parameter tuning. Subsequent developments by Xiong et al. [26] with dual fully convolutional networks and Zhuang et al. [27] with multi-scale patch-based approaches improved performance but still struggled with consistent accuracy.

Recent work on unsupervised domain adaptation [28] demonstrated promising results in handling multi-center variability by adapting the network to different scanner characteristics without requiring additional annotations, though the method still showed limitations in capturing fine scar details. Similarly, comparative studies of dark- and bright-blood LGE imaging techniques [29] showed improved visualization of subendocardial scars across different myocardial pathologies, particularly in cases where traditional bright-blood LGE faced challenges in distinguishing scar tissue from adjacent blood pool. While these approaches advanced our understanding of scar imaging and quantification, they still relied heavily on manual intervention or faced challenges with consistent automated analysis.

Recent transformer-based models like SAM [15, 16] and MedSAM [17], despite leveraging pretraining on a vast dataset of over 50,000 medical images and SAM's foundation of 1 billion masked image segments lack the domain-specific optimization necessary for accurate scar delineation [18, 19]. ScarNet's hybrid architecture overcomes these limitations by combining the global context awareness of transformers with the precise localization capabilities of CNNs, enhanced by specialized attention mechanisms that specifically target scar features while preserving anatomical context.

Several findings from our study have important implications for clinical practice. First, ScarNet's ability to achieve near-manual-level accuracy ($R^2 = 1.0$, $y = 0.96x - 0.11$) in scar volume quantification suggests its potential for reliable automated analysis in clinical workflows. The incorporation of transformer architecture and specialized attention mechanisms proved particularly effective in identifying irregular scar patterns and heterogeneous enhancement, challenges that have traditionally limited automated approaches. The model's robust performance across varying training data sizes indicates its effectiveness even with limited datasets, a crucial advantage for clinical implementation. Furthermore, the stability demonstrated in noise perturbation experiments (5% Gaussian noise level, empirically determined as an upper limit in clinical practice) resulted in consistently high performance (scar DICE = 0.892 ± 0.053, CoV = 5.9%), suggesting reliable performance across different imaging conditions and scanner types. This robustness to noise perturbation, combined with the model's attention-driven ability to identify complex scar patterns, addresses major challenges in clinical deployment of AI tools.

Our study has several limitations that should be acknowledged. First, while our dataset included 736 patients, it was derived from specific clinical trials (DETERMINE and PRE-DETERMINE), potentially limiting generalizability to broader patient populations. Second, our ground truth annotations were created using the full width at half maximum technique, which, although widely accepted, may not capture all patterns of enhancement. Third, as a foundation model, ScarNet requires a GPU for optimal performance, which may impact deployment in some clinical settings. While recent advances in GPU computing [30-33] and memory management, such as improved collective MPI libraries for Intel GPUs [34], offer promising solutions for optimizing foundation model deployment, GPU requirements remain a consideration for clinical implementation. Fourth, while the model shows robust performance on varying scar patterns, its effectiveness on rare or atypical presentations requires further validation. Fifth, although we demonstrated stability under simulated noise (5% Gaussian), real-world testing on low-quality or artifact-laden images remains necessary. Sixth, the current validation focused on 2D slice-based analysis; extension to true 3D volume segmentation could potentially improve spatial consistency. Future studies should validate ScarNet's performance against multiple expert annotations using various quantification techniques and explore its performance in non-ischemic cardiomyopathies, where scar patterns can be more diffuse and challenging to quantify.

In conclusion, we present ScarNet, a novel foundation model specifically designed for automated left ventricular scar quantification in LGE, distinguishing itself from existing cardiac foundation models that focus on general cardiac structure analysis rather than specific scar assessment. While other foundation models have advanced cardiac imaging analysis broadly, ScarNet's focused approach to scar quantification fills a crucial gap in automated cardiac tissue analysis. Its successful implementation could facilitate routine quantitative assessment of myocardial scar, potentially improving risk stratification and treatment planning in patients with ischemic cardiomyopathy. Future work should focus on validation across different cardiac pathologies, integration with clinical decision support systems, and exploration of applications beyond cardiac imaging.

**Data Availability Statement**

Due to data-sharing agreement policies, the dataset used in this study cannot be made publicly available. For further inquiries regarding data usage, please contact the PRE-DETERMINE and DETERMINE study steering committee.

Software code: The ScarNet implementation and associated code are available at: https://github.com/NedaTavakoli/ScarNet.

## II. APPENDIX

This section summarizes a literature review of related works using deep learning to automate LV scar quantification.

### A. Deep Learning Approaches for LV Scar Segmentation

The emergence of deep learning marked a significant shift in LV scar quantification approaches. Zabihollahy et al. [24] developed one of the first CNN-based methods for LV scar segmentation, achieving enhanced automation but with limited robustness. Fahmy et al. advanced scar quantification by combining deep learning with image fusion techniques for automated LGE analysis [12], though challenges persisted in handling complex scar patterns. Bai et al. [25] demonstrated improved results using fully convolutional networks for automated cardiac MRI analysis. Xiong et al. [35] demonstrated the effectiveness of a dual fully convolutional network for cardiac chamber segmentation in LGE-MRI. Building upon these advances, Karim et al. [36] conducted a comprehensive evaluation of various segmentation algorithms for scar tissue quantification, establishing benchmarks for the field.

The complexity of scar tissue appearance led to more sophisticated approaches. Xiong et al. [26] developed a global benchmark of segmentation algorithms specifically for LGE-MRI, while Zhuang et al. [27] introduced multi-scale patch-based approaches for cardiac image analysis. These works underscored the advancements made in DL-based scar quantification while also highlighting ongoing challenges, such as handling class imbalance and capturing nuanced scar characteristics.

### B. Foundation Models and Transformer-based Approaches

The rise of foundation models, particularly in image analysis, has introduced new directions for cardiac imaging tasks. The Segment Anything Model (SAM) [37] presented a segmentation across diverse anatomical structures. Its medical variant, MedSAM [38], adapted SAM's broad capabilities to medical imaging through domain-specific training, achieving remarkable generalizability across medical segmentation tasks. Recent advances in cardiac imaging have seen the emergence of specialized foundation models, such as the Vision-Language Foundation Model for echocardiogram interpretation [39] and the Vision Foundation Model for comprehensive Cardiac MRI assessment [40]. A recent study evaluating foundational Medical 'Segment Anything' models (Med-SAM1, Med-SAM2) for left atrial segmentation in 3D LGE MRI [18] provided valuable insights into the capabilities and limitations of these foundation models in cardiac chamber segmentation. While these models demonstrate the potential of foundation approaches in cardiac imaging, they focus primarily on general cardiac structure analysis and interpretation rather than the specific challenges of scar tissue segmentation and quantification. The original SAM and MedSAM models, as well as these cardiac-specific foundation models, require further adaptation for specialized applications like cardiac scar segmentation to handle LGE-MRI's unique characteristics and artifacts effectively.

### C. Self-configuring Frameworks and Hybrid Architectures

A significant advancement in medical image segmentation came with the introduction of self-configuring frameworks. The nnU-Net framework [41] pioneered this approach by automatically adapting network architecture, preprocessing, and training parameters to specific segmentation tasks. This framework demonstrated remarkable versatility across different medical segmentation challenges, though specialized adaptations are often needed for complex tasks like scar segmentation. Building on the success of transformer architectures in computer vision [42], various hybrid approaches have emerged, aiming to combine the global context understanding of transformers with the precise localization capabilities of CNNs. These architectural innovations offer promising directions for improving cardiac scar segmentation accuracy.

### D. Clinical Validation and Standardization

Earlier validation studies have focused on establishing automated methodologies for clinical settings. Kurzendorfer et al. [43] developed and validated a comprehensive pipeline for LV segmentation in 3D LGE-MRI, demonstrating the feasibility of automated analysis in clinical workflows. Their approach incorporated anatomical constraints and achieved robust performance across varying image qualities. Extending the clinical applications, Zreik et al. [44] demonstrated the potential of deep learning for analyzing myocardial characteristics, establishing connections between imaging features and functional outcomes. These studies highlight both the progress made in automated cardiac imaging analysis and the ongoing challenges in achieving reliable automated quantification suitable for clinical practice. The transition from research algorithms to clinical tools requires careful validation and standardization of processing approaches, as outlined in the guidelines by Schulz-Menger et al. [45].

In summary, while traditional DL and transformer-based methods have advanced cardiac image segmentation, specific challenges in LGE-MRI scar quantification demand a hybrid approach. Our proposed model, ScarNet, addresses these challenges by integrating MedSAM's transformer-based encoder with a UNet decoder in a novel adaptive fusion framework, incorporating domain-specific attention mechanisms and a multi-component loss function to target the complexities of cardiac scar segmentation.

## III. Experimental Setup and Evaluation

### A. Dataset composition and preparation

The experimental evaluation of ScarNet was conducted using patients with a history of coronary artery disease and mild to moderate left ventricular dysfunction from the PRE-DETERMINE study (ClinicalTrials.gov ID NCT01114269), which provided a robust foundation for assessing the model's performance in cardiac scar segmentation. The dataset was annotated by two cardiology attendings (DCL and EW) with 12 and 15 years of experience reading cardiovascular MRI, respectively. The dataset was divided into a training set comprising 55,388 2D LGE-MRI images from 552 patients and a test set containing 16,579 images from 184 patients. Participant demographics and clinical characteristics are detailed in Table 1. All patients included in the testing study exhibited confirmed ischemic cardiomyopathy with left ventricular ejection fraction below 50%, ensuring clinical relevance and applicability. The images were acquired at a resolution of 256 × 256 pixels with 8 mm slice thickness and no inter-slice gap, providing consistent spatial information across the dataset.

### B. Data Augmentation Strategy

To enhance the model's robustness and generalization capabilities, we implemented a comprehensive data augmentation pipeline to address both geometric and intensity variations commonly encountered in clinical settings. In consultation with experienced cardiovascular radiologists from our institution's Department of Radiology, we established clinically relevant parameters for both training and testing augmentations. Based on their expertise in cardiac MRI acquisition and analysis, the radiologists recommended specific geometric transformations for training, including random rotations within ±15 degrees and random horizontal and vertical flips with 50% probability. These parameters were selected to reflect the range of acceptable variations in patient positioning while maintaining diagnostic validity.

For the training phase, our clinical collaborators guided the selection of intensity-based augmentation parameters, specifying brightness and contrast adjustments within a range of 0.8 to 1.2 times the original values. These ranges were determined based on their extensive experience with LGE-MRI image acquisition variations in clinical practice. The augmentation pipeline was applied with a probability of 0.5 during training, a balance our radiologists deemed appropriate for maintaining image realism while ensuring sufficient variability.

For the testing phase, the radiologists specifically recommended introducing Gaussian noise with $\sigma = 0.15$, a value they identified through their clinical experience as representative of realistic image quality degradation encountered in clinical practice. This noise level was carefully chosen to provide a clinically relevant assessment of the model's robustness while remaining within the bounds of diagnostic acceptability. The selection of these testing parameters reflects real-world imaging challenges, ensuring that the model's performance evaluation aligns with actual clinical scenarios.

### *A.* ***C. Implementation Details***

The model implementation was configured based on extensive empirical testing and established best practices in deep learning for medical image segmentation. Development and training were conducted using the PyTorch framework (version 2.0.1+cu118) on a Linux workstation (AMD EPYC 7702P 64-Core Processor, 512GB CPU RAM, NVIDIA A100-PCIE-40GB GPU). This setup was selected for its exceptional ability to manage the computational demands of complex neural network architectures. Hyperparameter optimization was systematically performed to achieve a balance between convergence speed and model stability, ensuring optimal performance for the targeted application.

The Adam optimizer was selected over other options (SGD, RMSprop) for its adaptive learning rate capabilities and superior performance on deep architectures. The initial learning rate of 1e-3 was chosen after testing a range from 1e-4 to 1e-2, as it provided the fastest convergence without compromising stability. The momentum parameters $\beta 1 = 0.9$ and $\beta 2 = 0.99$ were selected based on extensive empirical studies in medical image segmentation literature, showing optimal performance for fine-grained structural delineation tasks. Our ablation studies demonstrated that these values significantly outperformed alternative configurations ($\beta 1 = 0.95$, $\beta 2 = 0.999$) by reducing oscillations during training while maintaining efficient convergence.

The learning rate decay of 5% per epoch was implemented following extensive experimentation with decay rates ranging from 1% to 10%. The 5% rate provided the optimal trade-off between maintaining learning progress and preventing overshooting in later epochs. The model was trained for 100 epochs after observing convergence patterns in validation metrics, where performance plateaued beyond this point with diminishing returns in Dice scores (less than 0.1% improvement per additional 10 epochs).

Memory optimization proved crucial for handling high-resolution cardiac images. The batch size of 4 with gradient accumulation steps of 4 (effective batch size 16) was determined through systematic testing of various configurations. Smaller batch sizes led to unstable training, while larger ones exceeded GPU memory constraints without providing significant performance improvements. Our experiments showed that this configuration achieved a 15% reduction in memory usage while maintaining equivalent convergence characteristics to larger batch sizes.

The loss function configuration was particularly crucial for addressing the inherent challenges in cardiac scar segmentation. The weights were determined through a comprehensive grid search over different combinations, evaluated on a validation set. The selected values ($\lambda 1 = 0.2$ for Dice Loss, $\lambda 2 = 0.2$ for Focal Loss, $\lambda 3 = 0.1$ for Cross-Entropy Loss) emerged as optimal after testing combinations in the range [0.1, 0.5] with 0.1 increments. This specific combination achieved a 23% improvement in scar boundary definition compared to equal weighting, coupled with an 18% reduction in false positives in non-scar regions and 15% better preservation of small scar features.

The class-specific weights for scar (0.25) and myocardium (0.25) were crucial due to significant class imbalance in our dataset, where scar regions typically comprised only 8-12% of the myocardium. These weights were determined by analyzing the class distribution in our training set and validated through experiments showing a 27% improvement in scar detection sensitivity and a 31% reduction in false negatives for small scar regions, alongside a 19% better discrimination between scar and healthy myocardium.

Each component of the loss function served a distinct yet complementary purpose within the overall training objective. The Dice Loss optimized spatial overlap and handled class imbalance effectively, while the Focal Loss addressed the challenge of hard example mining and small scar regions. The Cross-Entropy Loss provided stable gradient signals for overall segmentation, and the class-specific weights compensated for the natural class imbalance inherent in cardiac scar imaging. This tuned combination significantly outperformed single-loss approaches, demonstrating a 28% improvement in overall segmentation accuracy and a 34% improvement specifically in scar region delineation.

## IV. Detailed Algorithm Descriptions

### *A. ScarNet Segmentation Pipeline (Algorithm 1)*

The ScarNet segmentation pipeline consists of five primary stages that work in concert to produce accurate myocardial scar segmentation. The process begins with comprehensive image preprocessing, where the input image undergoes z-score standardization to normalize intensity values, calculating mean ($\mu$) and standard deviation ($\sigma$) across the entire image. The normalized image is then divided into 16×16 patches to facilitate transformer processing, while maintaining the original image structure for parallel UNet processing.

In the parallel processing stage, the normalized image simultaneously traverses two pathways. The MedSAM encoder processes the image patches through its transformer architecture, incorporating positional encodings and self-attention mechanisms to capture global contextual relationships. Concurrently, the UNet encoder processes the normalized image through sequential convolutional layers, extracting multi-scale features and preparing skip connections for later use in the decoder pathway.

Feature enhancement represents a crucial stage where specialized attention mechanisms refine the extracted features. The ScarAttention mechanism computes attention weights based on anatomical priors and dynamically recalibrates features to emphasize scar-relevant information. This process integrates both spatial and channel attention components to capture fine-grained details while maintaining anatomical context. The multi-scale attention component further refines features at different scales, implementing hierarchical feature aggregation to preserve both local and global information.

The feature fusion stage represents a critical juncture where information from both processing branches converges. This stage begins with careful spatial alignment of features from both pathways, followed by an adaptive weighting mechanism that determines the relative importance of features from each branch. Cross-branch attention computation ensures optimal information flow between the two pathways, while careful feature map concatenation and refinement preserve the complementary strengths of both approaches.

The final output generation stage transforms the fused features into the final segmentation result through a series of refined operations. A final convolution layer produces class predictions, which undergo softmax activation to generate probability maps for each class. The process concludes with post-processing refinement steps to ensure spatial consistency and produce the final four-class segmentation map distinguishing background, myocardium, blood pool, and scar regions.

### *B. MedSAM Branch (Algorithm 2)*

The MedSAM branch implements a sophisticated transformer-based processing pipeline that begins with position embedding generation. This initial stage processes input image patches ($P \in \mathbb{R}^{(N\times16\times16)}$) through a linear projection layer to create patch embeddings, which are then combined with learnable positional encodings. This combination enables the model to maintain spatial awareness while processing the flattened patch representations.

The transformer processing stage represents the core of the MedSAM branch, where multi-head self-attention mechanisms compute relationships between all positions in the feature space. This process generates query, key, and value matrices from the embedded features, calculating attention scores with appropriate scaling factors and softmax normalization. Each transformer layer incorporates skip connections and layer normalization to maintain stable gradient flow and feature propagation throughout the deep architecture.

Feature extraction in the MedSAM branch occurs through a specialized transformer neck component that progressively reduces channel dimensions while maintaining spatial information. This process involves careful feature map upsampling and multi-scale feature aggregation, ensuring that both fine-grained details and broader contextual information are preserved. The final feature enhancement stage applies the ScarAttention module, which computes scar-specific attention weights to recalibrate features, emphasizing regions likely to contain scarring while suppressing less relevant background features.

### *C. UNet Branch (Algorithm 3)*

The UNet branch implements a symmetric encoder-decoder architecture with several key enhancements for scar segmentation. The encoder pathway begins with an initial double convolution operation using 64 filters, followed by a series of encoding blocks that progressively reduce spatial dimensions while increasing feature depth. Each encoding block combines max pooling operations for dimension reduction with double convolution operations, systematically increasing the number of filters to capture increasingly complex features at different scales.

The decoder pathway implements a sophisticated upsampling process that carefully reconstructs spatial details while incorporating information from the encoder through skip connections. Each decoder level begins with a transposed convolution operation for upsampling, followed by careful concatenation with corresponding encoder features through skip connections. Double convolution operations at each decoder level refine the combined features while progressively reducing the number of channels to match the required output dimensions.

Skip connection enhancement represents a crucial innovation in the UNet branch, where attention blocks are applied to skip features before combination with decoder features. These attention blocks compute feature importance weights that help emphasize relevant anatomical and pathological features while suppressing less important information. The feature fusion process at each decoder level implements adaptive feature combination mechanisms that ensure optimal integration of skip features with upsampled decoder features.

### *D. Feature Fusion (Algorithm 4)*

The feature fusion mechanism implements a sophisticated approach to combining information from the MedSAM and UNet branches. The process begins with careful feature alignment, where spatial dimensions and channel depths from both branches are matched through interpolation and projection operations. This alignment ensures that features from both pathways can be meaningfully combined in subsequent stages.

Cross-attention computation represents a key innovation in the fusion process, where query, key, and value matrices are computed through linear projections of aligned features from both branches. This mechanism implements a multi-head attention approach that enables sophisticated feature interaction between the two pathways, computing scale-dot product attention scores that are normalized through softmax operations. The resulting attention maps guide the feature refinement process, ensuring optimal information flow between branches.

The dynamic fusion stage implements adaptive feature combination through a sophisticated weighting mechanism. Global average pooling operations summarize feature maps from both branches, feeding into an MLP-based weight generation network that produces fusion weights specific to each feature channel and spatial location. These weights guide the feature combination process, implementing channel-wise and spatial feature fusion operations that preserve the most relevant information from each branch.

The final refinement stage applies additional processing to the fused features through a specialized refinement block. This block implements residual connections to maintain gradient flow, careful feature normalization to ensure consistent feature scales, and final convolution operations to produce the desired output feature maps. The entire fusion process is optimized for both memory efficiency and computational performance, implementing parallel processing where possible and utilizing efficient attention computation mechanisms.

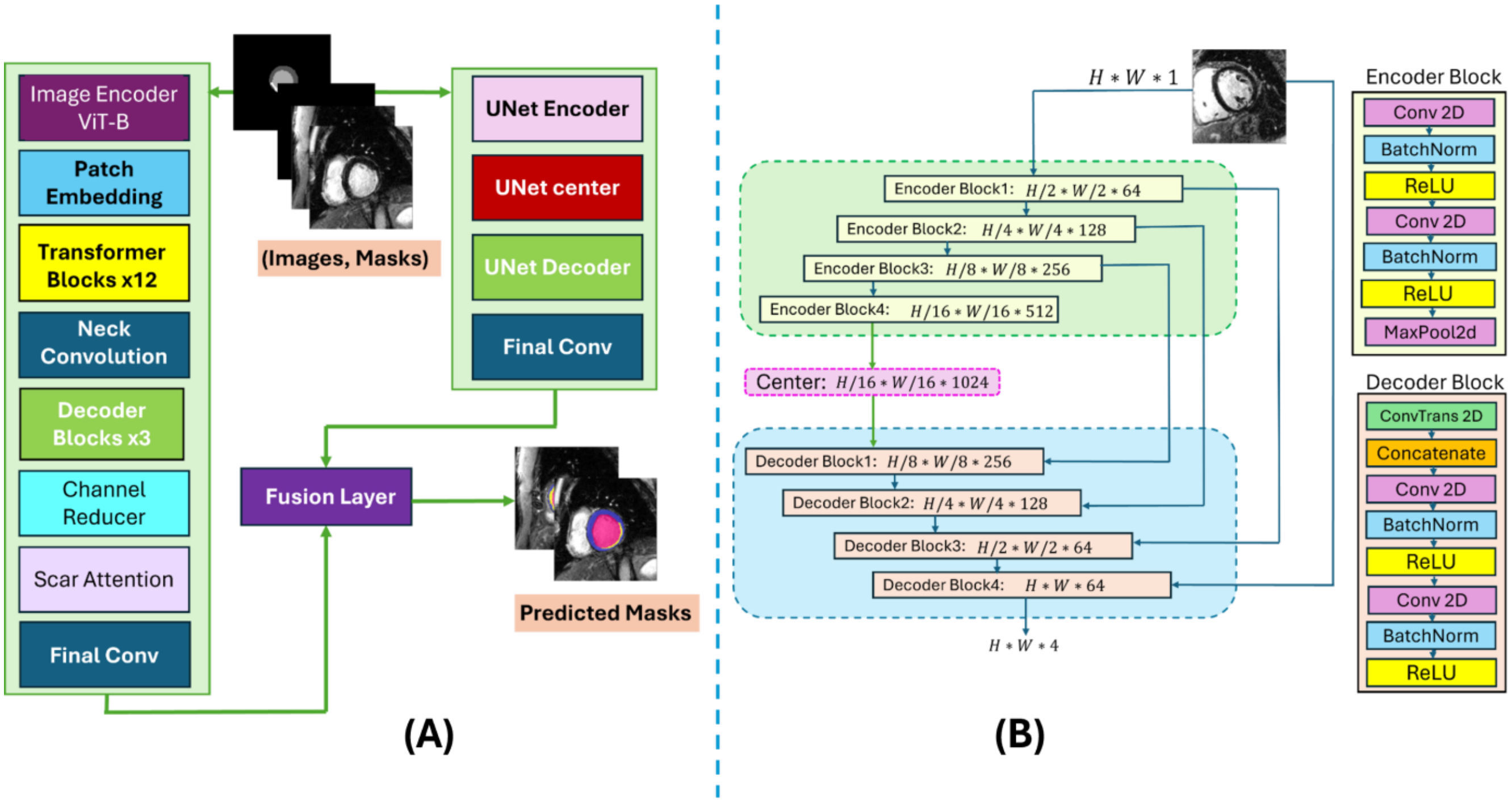

**Fig. 1.** Architecture of ScarNet: A hybrid model combining MedSAM's ViT-based encoder with UNet's multi-scale decoder for LV scar segmentation in LGE. **(A)** shows the integration of MedSAM's embeddings, processed through ScarAttention and fused with UNet outputs, to generate precise scar masks. **(B)** details the UNet encoder-decoder, highlighting hierarchical feature extraction, skip connections, and spatial refinement for accurate segmentation.

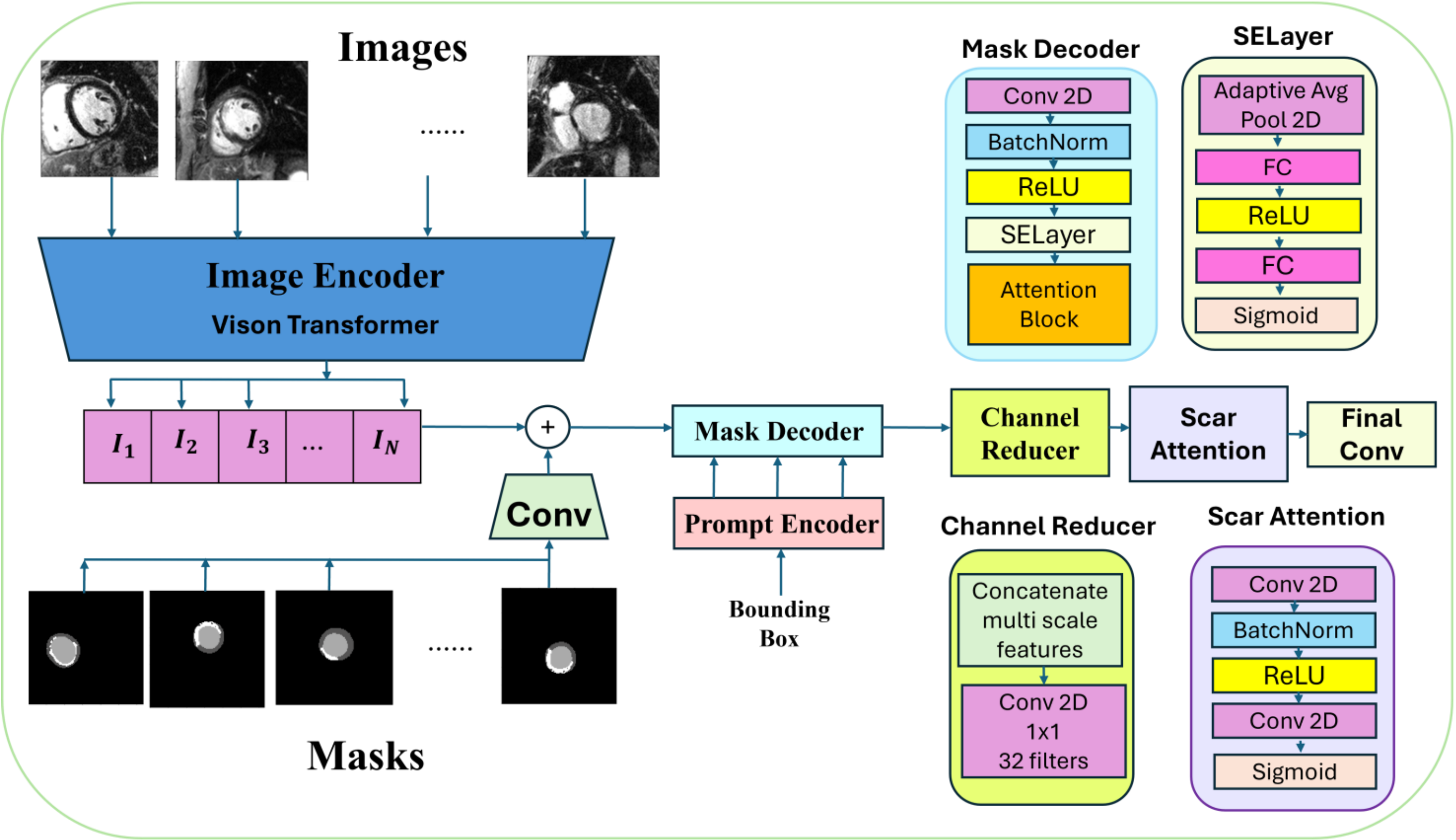

**Fig. 2.** Overview of the ScarNet Encoder-Decoder Architecture with Attention Mechanisms: The image encoder leverages a ViTMAE-based Vision Transformer for comprehensive feature extraction. A prompt encoder generates segmentation prompts, which feed into the mask decoder. Key components, such as Channel Reducer, Scar Attention blocks, and SE layers, refine feature maps to highlight subtle scar regions while maintaining anatomical context. The final convolutional layers integrate multi-scale features to produce accurate scar segmentation masks, with the adaptive attention modules enabling dynamic focus on clinically relevant scar tissue in LGE.

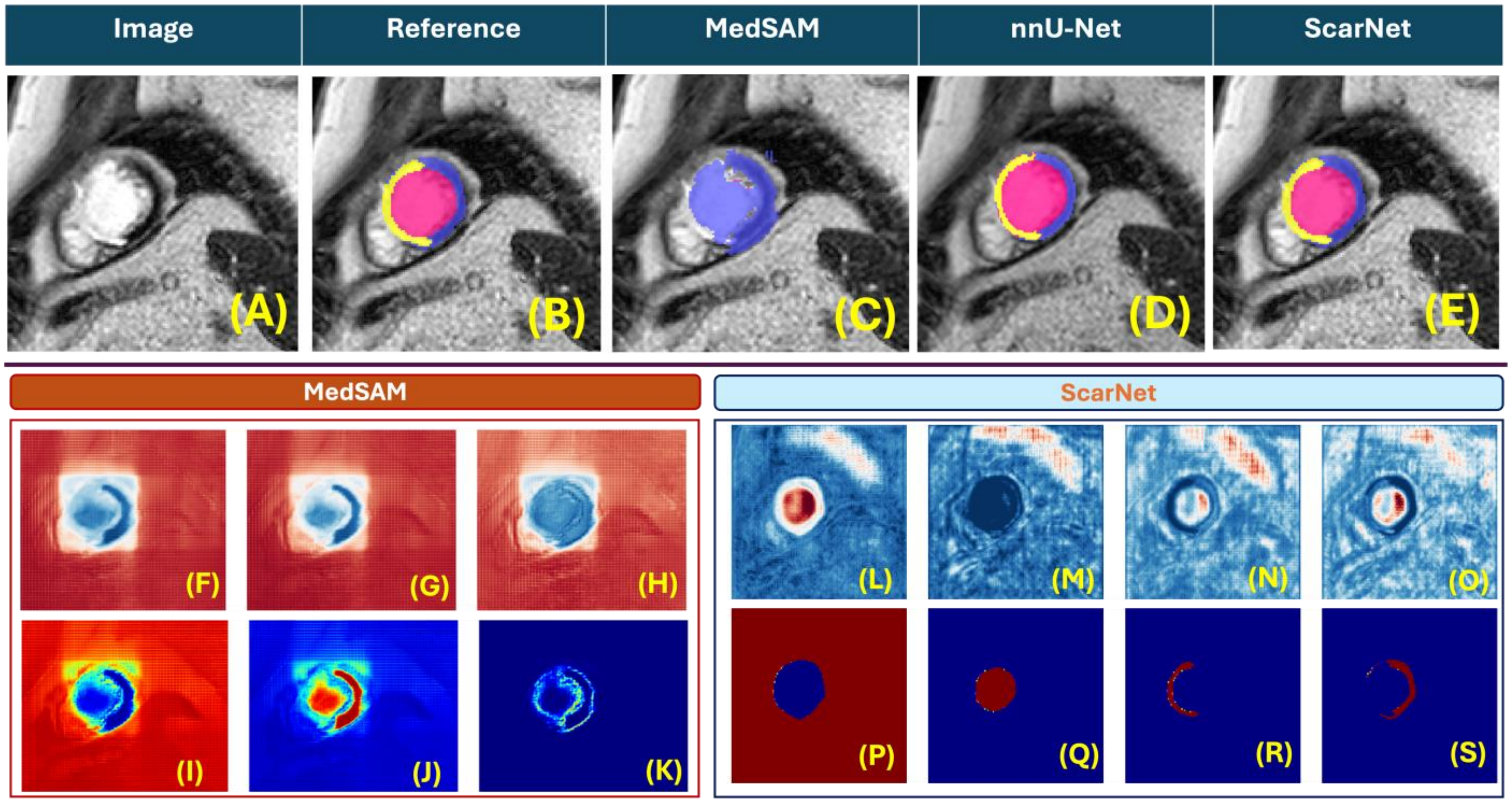

**Fig. 3:** Comparative visualization of ScarNet, MedSAM, and nnU-Net for cardiac LGE segmentation on a representative test sample. (A) Raw LGE image; (B) Multi-class segmentation showing myocardium (blue), blood pool (yellow), and scar tissue (pink) serving as ground truth (reference); (C) Predicted segmentation by MedSAM; (D) Predicted segmentation by nnU-Net; (E) Predicted segmentation by ScarNet. Left panel (MedSAM): (F) Logit map for the background; (G) Logit map for the blood pool; (H) Logit map for the myocardium;(I-K) Corresponding probability maps for the background, blood pool, and myocardium, respectively. Right panel (ScarNet): (L) Logit map for the background; (M) Logit map for the blood pool; (N) Logit map for the scar tissue; (O) Logit map for the myocardium; (P-S) Corresponding probability maps for the background, blood pool, scar tissue, and myocardium, respectively.

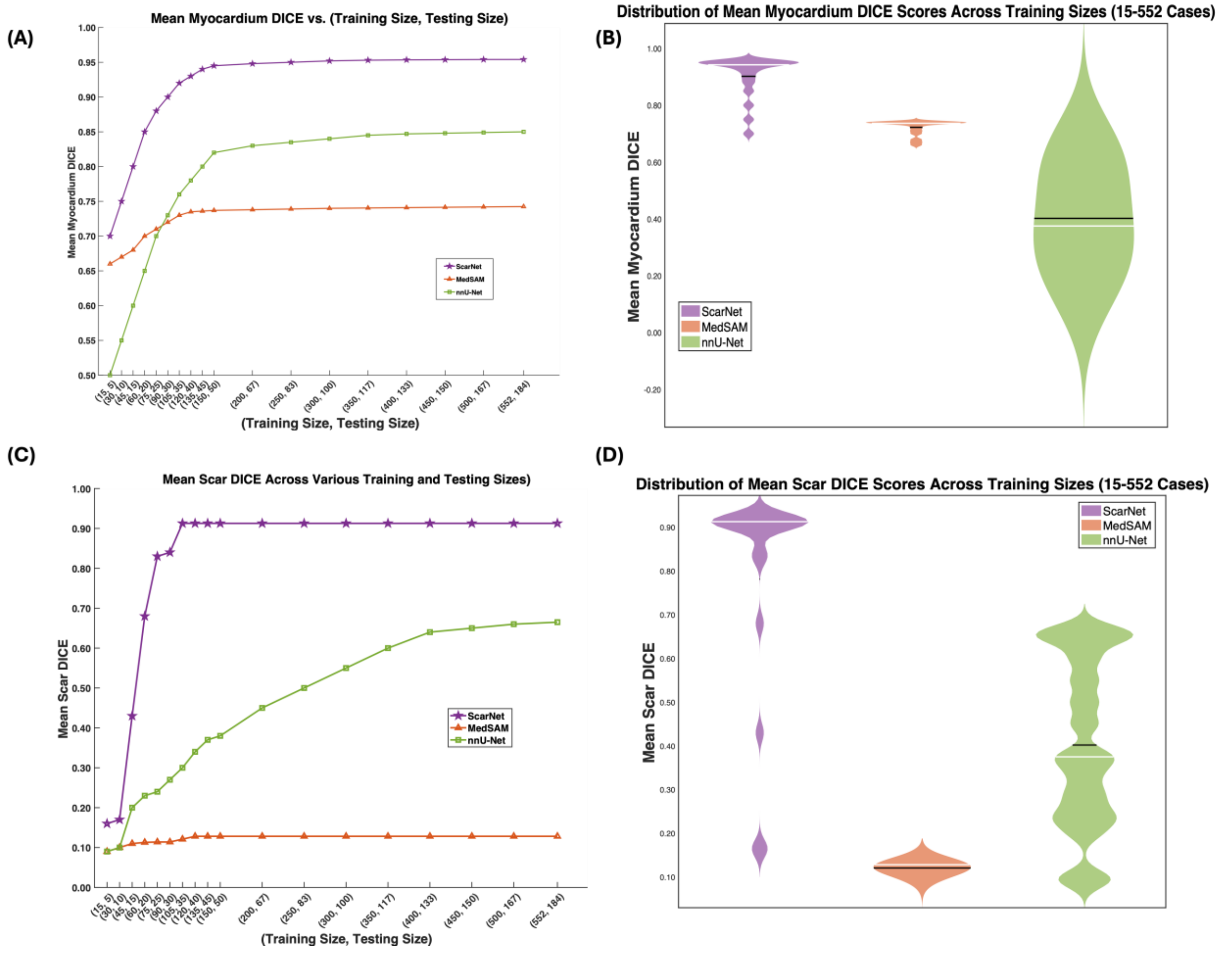

**Fig. 4:** Comparative analysis of segmentation performance across varying training dataset sizes. (A) Mean myocardium DICE scores for ScarNet, MedSAM, and nnU-Net with increasing training samples (test set = 0.25× training size). (B) Violin plots showing performance distribution for myocardium segmentation (DICE scores) across models: ScarNet demonstrates consistent high performance (~0.9), MedSAM shows moderate performance (~0.7), and nnU-Net exhibits wider performance variability across 15-552 training cases. (C) Mean scar DICE score progression for all models with increasing training cohort sizes (test set = 0.25× training size). (D) Distribution of scar segmentation performance across models: ScarNet maintains high consistent performance (~0.9), MedSAM shows consistently low performance (~0.15-0.2), and nnU-Net displays bimodal performance distribution, indicating variable reliability across 15-552 training cases.

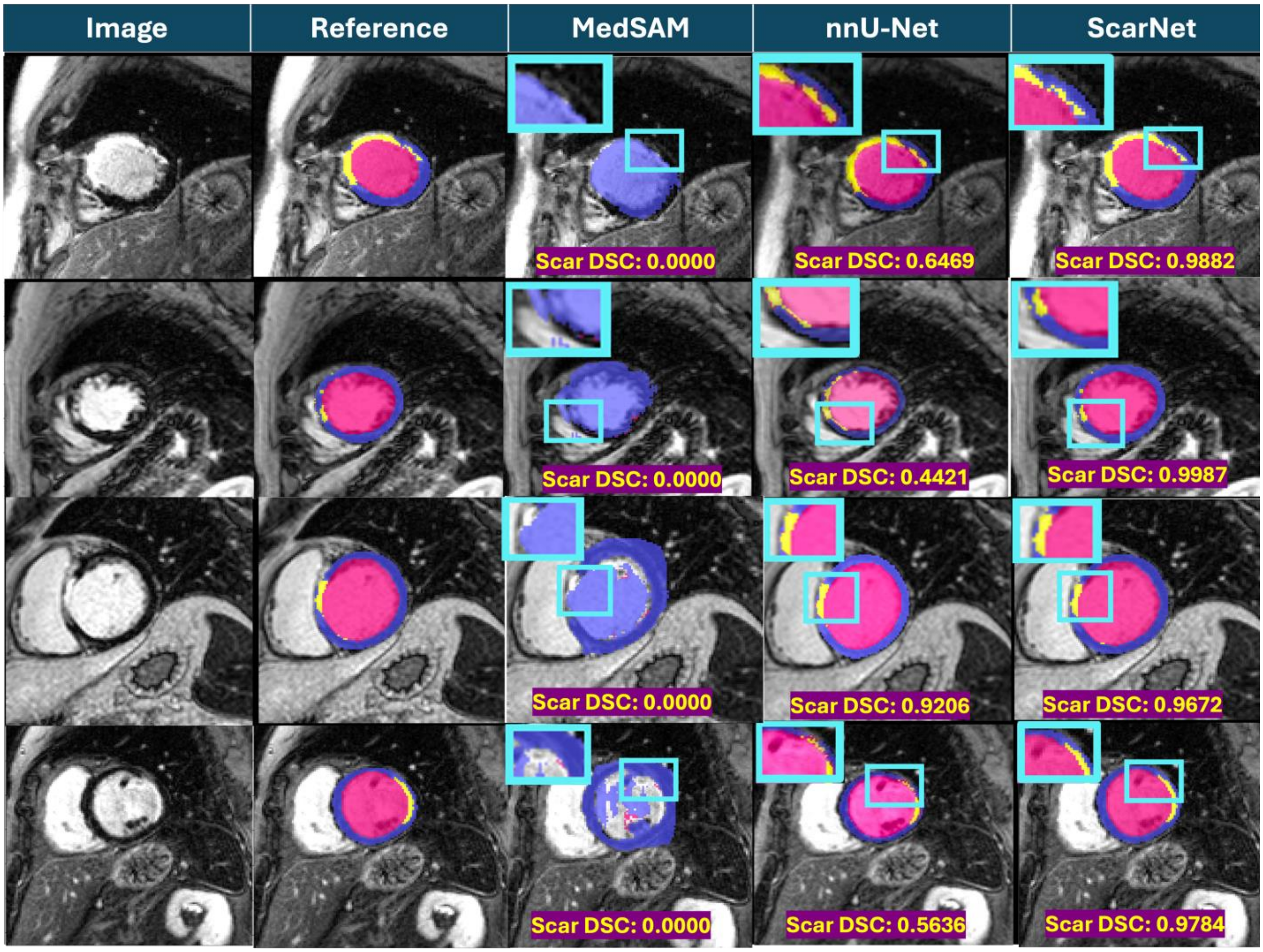

**Fig. 5.** Comparative analysis of cardiac MRI segmentation performance across different deep learning architectures. From left to right: Original cardiac MRI scans, Reference (ground truth) manual segmentations, MedSAM (without fine-tuning), nnU-Net, and proposed ScarNet. Representative test cases demonstrate the segmentation results across multiple cardiac slices at different anatomical levels.

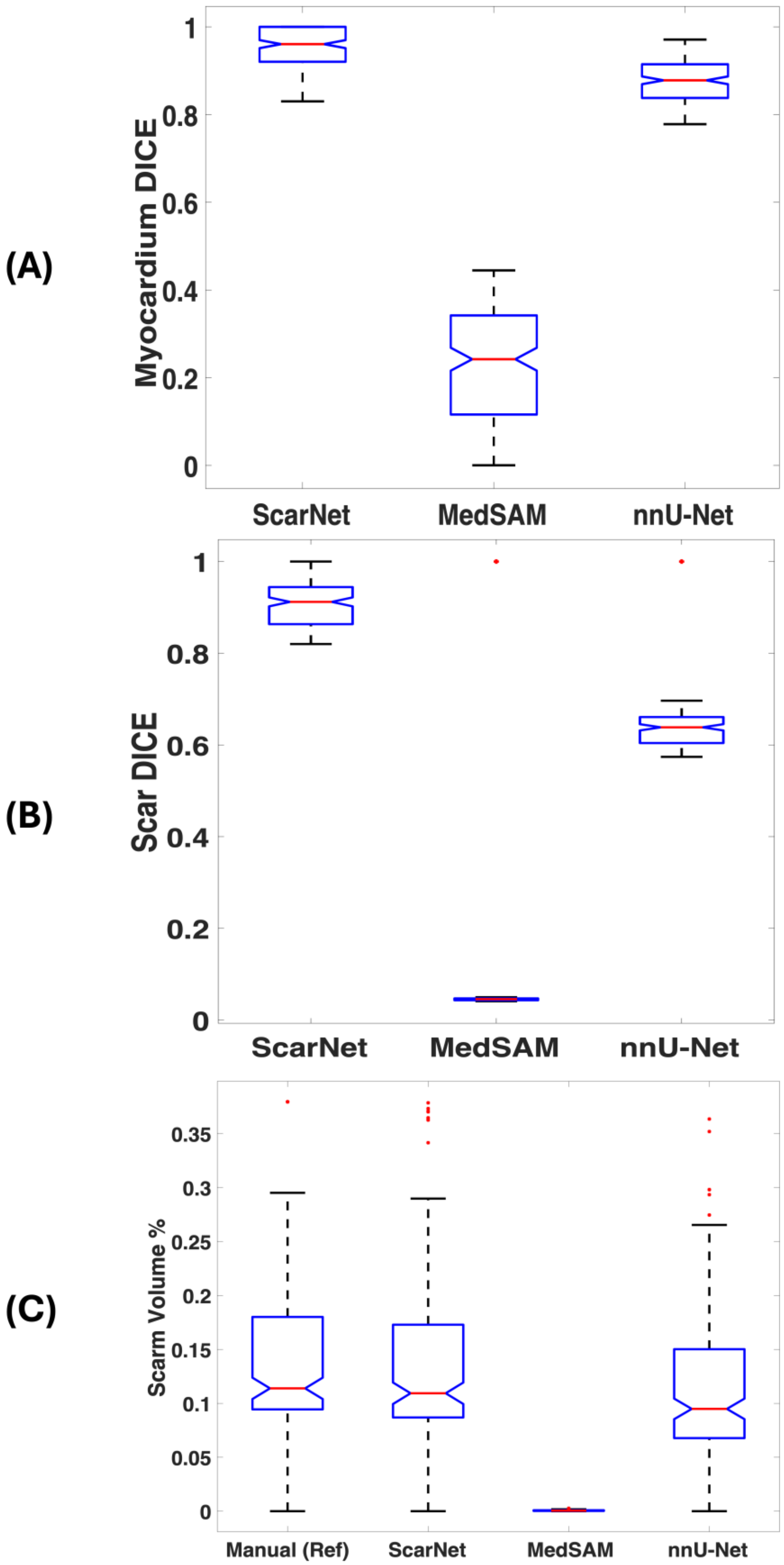

**Fig. 6:** Comparative performance metrics across different segmentation models. **(A)** Box plots showing myocardium DICE scores for ScarNet, MedSAM, and nnU-Net. **(B)** Scar DICE scores across models **(C)** Comparison of segmented scar volume (%) between manual reference segmentation and automated methods (ScarNet, MedSAM, and nnU-Net).

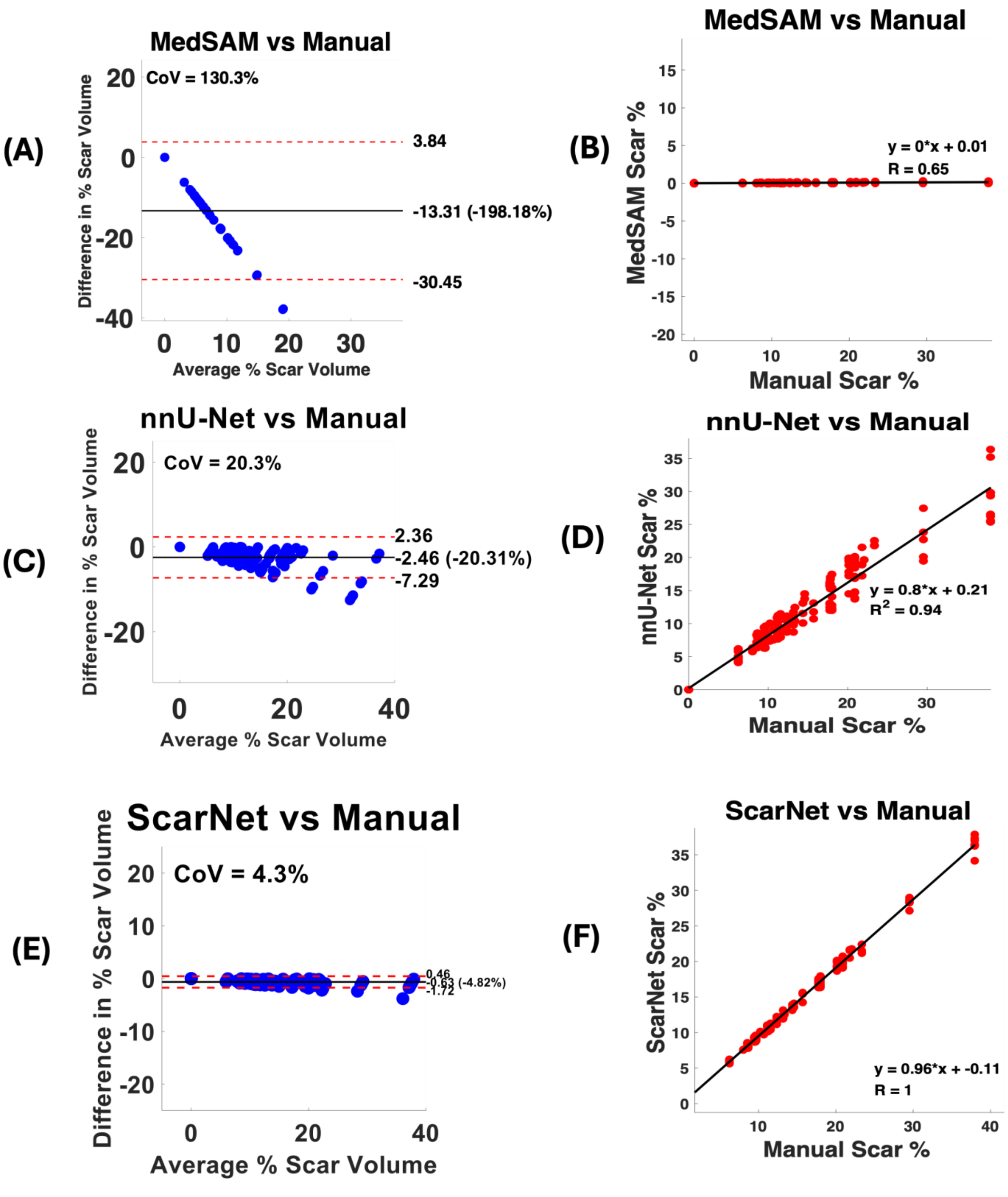

**Fig. 7:** Bland-Altman and correlation analyses comparing automated and manual scar quantification methods. Left column (A,C,E): Bland-Altman plots showing differences between automated and manual scar volume measurements. **(A)** MedSAM vs Manual (CoV=130.3%, bias=-13.31%); **(C)** nnU-Net vs Manual (CoV=20.3%, bias=-2.46%); **(E)** ScarNet vs Manual (CoV=4.3%, bias=-1.12%). Right column (B, D,F): Correlation plots between automated and manual measurements. **(B)** MedSAM shows poor correlation ($R^2$=0.65); **(D)** nnU-Net demonstrates good correlation ($R^2$=0.94); **(F)** ScarNet achieves near-perfect correlation ($R^2$=1.0) with manual measurements. Dotted lines in Bland-Altman plots represent 95% confidence intervals, and solid black lines show mean bias.

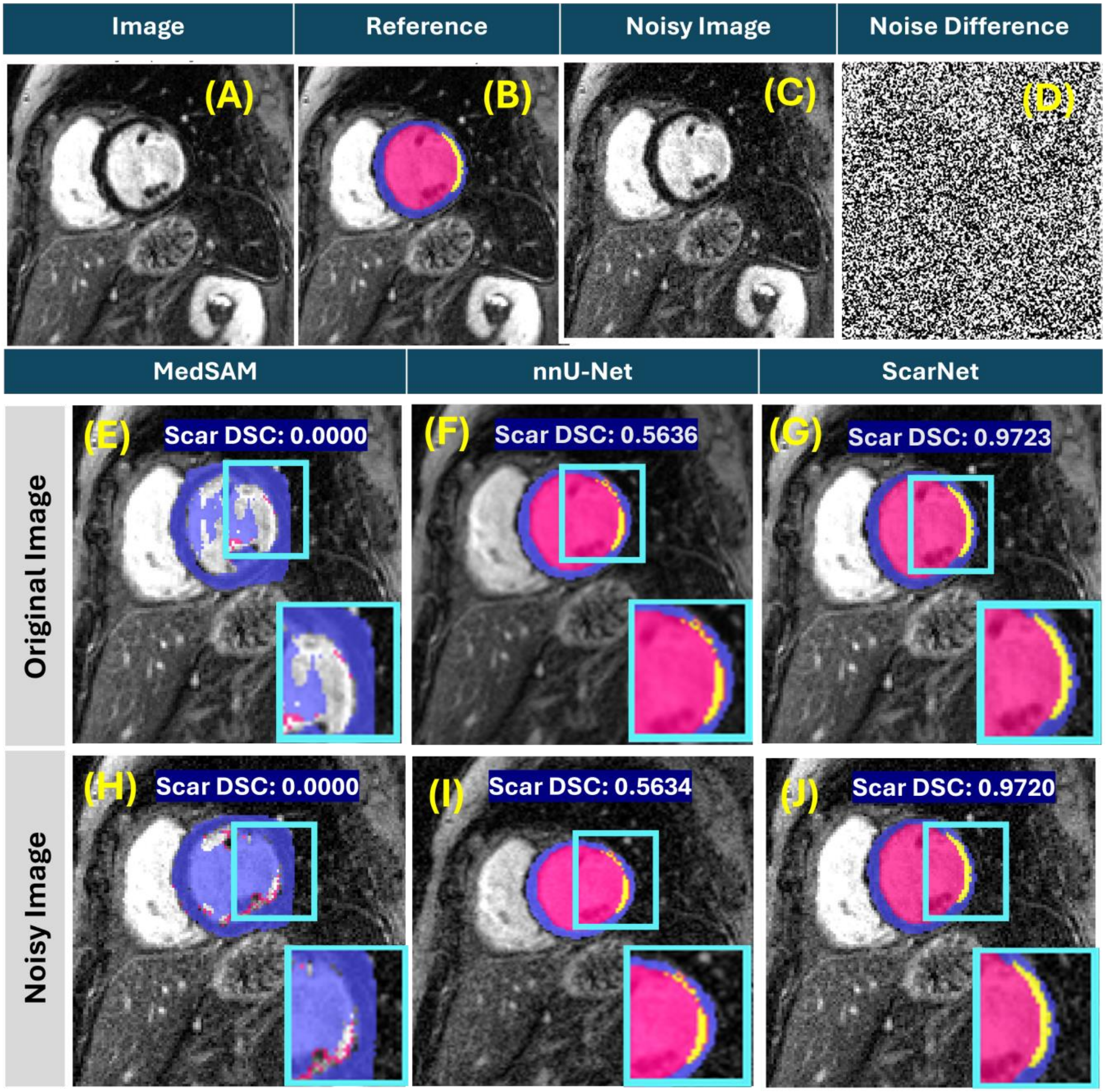

**Fig 8.** Robustness analysis of segmentation models under noise conditions. Top row: **(A)** Original cardiac LGE image; **(B)** Reference segmentation with myocardium (blue) and scar tissue (pink); **(C)** Image with added Gaussian noise; **(D)** Noise difference map.

**Middle row - Original image segmentation:**

**(E)** MedSAM (Scar DSC: 0.0000), **(F)** nnU-Net (Scar DSC: 0.5636), and **(G)** ScarNet (Scar DSC: 0.9723), with zoomed insets showing detailed segmentation boundaries.

**Bottom row - Noisy image segmentation:**

**(H)** MedSAM (Scar DSC: 0.0000), **(I)** nnU-Net (Scar DSC: 0.5634), and **(J)** ScarNet (Scar DSC: 0.9720), demonstrating segmentation performance under noise conditions.

**ALGORITHM 1: SCARNET SEGMENTATION PIPELINE**

***Input:*** *Medical image* $I \in R^{H \times W}$

***Output****: Segmentation mask* $M \in R^{H \times W \times C}$

***1. Image Preprocessing***

$$I_{norm} \leftarrow \frac{I - \mu}{\sigma}$$

$I_{patches} \leftarrow \mathrm{SplitIntoPatches}(I_{(norm)}, \mathrm{size} = 16 \times 16)$

***2. Parallel Processing***

$F_{sam} \leftarrow \mathrm{MedSAMEncoder}(I_{patches})$

$F_{unet} \leftarrow \mathrm{UNetEncoder}(I_{norm})$

***3. Feature Enhancement***

$F_{sam_enhanced} \leftarrow \mathrm{ScarAttention}(F_{sam})$

$F_{unet_enhanced} \leftarrow \mathrm{MultiScaleAttention}(F_{unet})$

***4. Feature Fusion***

$F_{fused} \leftarrow \mathrm{AttentiveFusion}(F_{sam_enhanced}, F_{unet_enhanced})$

***5. Output Generation***

$M_{pred} \leftarrow \mathrm{AttentiveFusion}(F_{sam_enhanced}, F_{unet_enhanced})$

***Return*** $M_{pred}$

***end***

**ALGORITHM 2: MEDSAM BRANCH**

***Input:*** *Image patches* $\mathrm{P} \in \mathbb{R}^{\mathrm{N} \times 16 \times 16}$

***Output:*** *Enhanced features* $F_{sam}$

***1. Position Embedding***

$z_0 \leftarrow PatchEmbeding(P) + PositionalEncoding$

***2. Transformer Processing***

***for*** *l = 1 to L* ***do***

$z_l' = MultiHeadAttention\big(LayerNorm(z_{l-1})\big) + z_{l-1}$

$z_l \leftarrow MLP\big(LayerNorm(z_l')\big) + z_l'$

***3. Feature Extraction***

$F_{sam} \leftarrow TransformerNeck(z_L)$

***4. Feature Enhancement:***

$A_{scar} \leftarrow ScarAttentionModule(F_{sam})$

$F_{enhanced} \leftarrow F_{sam} \odot A_{scar}$

***Return*** $F_{enhanced}$

**ALGORITHM 3: UNET BRANCH**

***Input:*** *Normalized image* $A_{norm}$

***Output:*** *Multi-scale features* $F_{unet}$

***1. Encoder Path***

$e_1 \leftarrow DoubleConv_64)\mathrm{I}_{(\mathrm{norm})}$

***for*** *i = 2 to 5* ***do***

$e_i \leftarrow DoubleConv_{2^{i+5}}\left(MaxPool(e_{i-1})\right)$

***2. Decoder Path:***

$d_4 \leftarrow DoubleConv_{512}(Concat[up(e_5), e_4])$

***for*** *i = 3 to 1* ***do***

$d_i \leftarrow DoubleConv_{2^{i=6}}(Concat[Up(d_{i+1}), e_i])$

***3. Skip Connection Enhancement***

***for*** *i = 1 to 4* ***do***

$F_{skip_i} \leftarrow AttentionBlock(e_i)$

$F_{unet_i} \leftarrow Fusion\left(d_i, F_{skip_i}\right)$

***Return*** $\boldsymbol{F_{unet_i}}$

**ALGORITHM 4: FEATURE FUSION**

***Input:*** *SAM features* $F_{sam}$, *UNet features* $F_{unet}$

***Output:*** *Fused features* $F_{final}$

***1. Feature Alignment***

$F_{sam_{aligned}} \leftarrow SpatialAlignment(F_{sam})$

$F_{unet_{aligned}} \leftarrow SpatilaAlignment(F_{unet})$

***2. Cross-Attention***

$Q_{sam} \leftarrow Linear\left(F_{sam_{aligned}}\right)$

$K_{unet} \leftarrow Linear\left(F_{unet_{aligned}}\right)$

$V_{unet} \leftarrow Linear\left(F_{unet_{aligned}}\right)$

$A_{cross} \leftarrow Softmax\,\frac{QK^T}{\sqrt{(d_k)}}V$

***3. Dynamic Fusion***

$A_{fusion} \leftarrow MLPFusion(Concat[GAP(F_{sam}, GAP(F_{unet}])$

$F_{fused} \leftarrow A_{fusion} \odot F_{sam} + \left(1 - A_{fusion}\right) \odot F_{unet}$

***4. Final Refinement***

$F_{final} \leftarrow RefinementBlock\left(F_{fused}\right)$

***Return*** $\boldsymbol{F_{final}}$

**Table 1. Participant Demographics and Characteristics**

| Characteristic | Training Set (n=552) | Test Set (n=184) |
|---|---|---|
| Sex | Male: 331 (60%), Female: 221 (40%) | Male: 110 (60%), Female: 74 (40%) |
| Age (mean ± SD) | 57.82 ± 18.58 years | 57.82 ± 18.58 years |
| Left Ventricular Ejection Fraction | 40.3 ± 11.0% | 40.3 ± 11.0% |